\documentclass[aip,graphicx]{revtex4-1}

\usepackage{graphicx}

\begin{document}


\title{Carbon-carbon supercapacitors: Beyond the average pore size or how electrolyte confinement and inaccessible pores affect the capacitance} 

\author{El Hassane Lahrar}
\affiliation{CIRIMAT, Universit\'e de Toulouse, CNRS, B\^at. CIRIMAT, 118, route de Narbonne 31062 Toulouse cedex 9, France}
\affiliation{R\'eseau sur le Stockage \'Electrochimique de l'\'Energie (RS2E), F\'ed\'eration de Recherche CNRS 3459, HUB de l'\'Energie, Rue Baudelocque,  80039 Amiens, France}

\author{Patrice Simon}
\affiliation{CIRIMAT, Universit\'e de Toulouse, CNRS, B\^at. CIRIMAT, 118, route de Narbonne 31062 Toulouse cedex 9, France}
\affiliation{R\'eseau sur le Stockage \'Electrochimique de l'\'Energie (RS2E), F\'ed\'eration de Recherche CNRS 3459, HUB de l'\'Energie, Rue Baudelocque,  80039 Amiens, France}

\author{Céline Merlet}
\email{merlet@chimie.ups-tlse.fr}
\affiliation{CIRIMAT, Universit\'e de Toulouse, CNRS, B\^at. CIRIMAT, 118, route de Narbonne 31062 Toulouse cedex 9, France}
\affiliation{R\'eseau sur le Stockage \'Electrochimique de l'\'Energie (RS2E), F\'ed\'eration de Recherche CNRS 3459, HUB de l'\'Energie, Rue Baudelocque,  80039 Amiens, France}

\date{\today}

\begin{abstract}
Carbon-carbon supercapacitors are high power electrochemical energy storage systems which store energy through reversible ion adsorption at the electrode-electrolyte interface. Due to the complex structure of the porous carbons used as electrodes, extracting structure-property relationships in these systems remains a challenge. In this work, we conduct molecular simulations of two model supercapacitors based on nanoporous electrodes with the same average pore size, a property often used when comparing porous materials, but different morphologies. We show that the carbon with the more ordered structure, and a well defined pore size, has a much higher capacitance than the carbon with the more disordered structure, and a broader pore size distribution. We analyze the structure of the confined electrolyte and show that the ions adsorbed in the ordered carbon are present in larger quantities and are also more confined than for the disordered carbon. Both aspects favor a better charge separation and thus a larger capacitance. In addition, the disordered electrodes contain a significant amount of carbon atoms which are never in contact with the electrolyte, carry a close to zero charge and are thus not involved in the charge storage. The total quantities of adsorbed ions and degrees of confinement do not change much with the applied potential and as such, this work opens the door to computationally tractable screening strategies.
\end{abstract}

\maketitle

\section{Introduction}

Electrochemical double layer capacitors (EDLCs), also known as supercapacitors, are energy storage systems attracting attention for their very high power densities, their wide range of operating temperatures and the large number of cycles of charge/discharge they can sustain.~\citep{Zhong15,Beguin14,Brandt13,Lewandowski10} In these systems, unlike for batteries, the energy is stored through reversible ion adsorption at the electrode surface, without faradaic reactions. As a consequence, supercapacitors are characterized by a relatively low energy density compared to batteries. As the energy stored in a supercapacitor is proportional to the capacitance and the square of the operating voltage, improvements will come from optimizing the electrode/electrolyte combination, by changing the electrode material or the electrolyte nature. The electrolyte impacts both the electrochemical window, i.e. the range of voltages for which the electrolyte does not decompose, and the capacitance. However, constraints, such as safety concerns and ionic conductivity, limit the number of suitable electrolytes.~\citep{Pal,Cai,Zhu} Carbon based materials, commonly used as electrodes in EDLCs as a result of their relatively cheap cost and correct electronic conductivity, can be synthesized with a wide range of morphologies which renders it possible to explore the relationships between electrode material properties and electrochemical performance.~\citep{Zhang2,Simon1,Gu3}

In idealized parallel plate capacitors consisting of metalling electrodes enclosing a dielectric medium, the capacitance is proportional to the surface area of the electrodes and inversely proportional to the distance separating the two electrodes (which could be approximated by the ion-carbon distance in supercapacitors). The use of porous carbon electrodes was thus envisioned early on as such materials can have large accessible surfaces. It should be noted though that while the presence of pores systematically increases the specific surface, it does not necessarily increase the surface accessible to the electrolyte. This has been shown, for example, by Lin~\emph{et~al.}~\citep{Lin99} who studied carbons with different pore sizes (micropores and mesopores) produced via a sol-gel process. They have shown that the micropores are not accessible to ions and do not contribute to capacitance. Endo~\emph{et~al}.~\citep{Endo01} arrived at the same conclusion using a variety of activated carbons. Following these results and those of similar works~\citep{Cazorla,Sing,Yoshida}, microporous materials were neglected for a time.

In 2006, a number of studies challenged the hypothesis that ions could not enter the micropores. Frackowiak~\emph{et~al.}~\citep{Frackowiak06} synthesized a series of template carbons from mesoporous silica and showed the existence of a linear correlation between the capacitance and the volume of the micropores. In these materials, it seems that the ultramicropores, with a size less than 0.7~nm, participate in the formation of the electrochemical double layer, provided that they are in the proximity of mesopores, with a size of 3~nm or more. Wang~\emph{et~al.}~\citep{Wang08} also observed the necessity for the interconnection of pores of different sizes so that the electrolyte can enter the micropores. At the same time, Chmiola~\emph{et~al.}~\citep{Chmiola06} tested Carbide Derived Carbons (CDC) as electrode materials for supercapacitors with sulfuric acid H$_2$SO$_4$ as electrolyte. These porous carbons are good model materials because of their narrow and controllable pore size distribution. The authors showed that the capacitance closely reflects the increase in microporous surface area. In addition, for these carbons, the increase in the porous volume for pores smaller than 2~nm increases the specific capacitance, while the increase in the porous volume for pores larger than 2~nm has a negative effect on capacitance. This shows that micropores not only participate in the construction of the electrical double layer, but they are also beneficial. Another study by Chmiola~\emph{et~al.}~\citep{Chmiola06b} on TiC-CDC in the presence of [NEt$_4$][BF$_4$] in acetonitrile showed that the pores with a size less than 1~nm participate largely in the capacitance. In fact, CDC allowed to reach previously unmatched capacitances of 140~F~g$^{-1}$. A remarkable capacitance increase was also measured for more common activated carbons,~\citep{Raymundo-Pinero06} which shows the generality of the performance of nanopores for energy storage. This result has generated a lot of research to understand the mechanisms of charge storage within nanopores, and to understand the structural and dynamic properties of confined electrolytes.~\citep{Salanne16,Forse16,Beguin14,Miao20}

Following the experimental observations of a large capacitance increase for pore sizes smaller than 1~nm,~\citep{Chmiola06b,Chmiola08,Raymundo-Pinero06}, many theoretical studies have been carried out to reproduce and understand this phenomenon. Molecular simulations have been conducted on a number of model supercapacitors with various electrode geometries going from planar electrodes~\citep{Kislenko09,Feng10b,Shim11,Shim12,Vatamanu10b,Feng11}, slit pores~\citep{Kondrat14,Kondrat16,Feng11,Lynden-Bell07,Wang19,Futamura17} or carbon nanotubes~\citep{Shim10,Frolov12,Yang09,Singh10,Vatamanu10} to more complex carbons~\citep{Mendez-Morales18,Soolo12,Merlet12,Liu19,Merlet13d}. Slit pores provide a good model geometry to study systematically the pore size dependence of the capacitance. Among the physical insights acquired through molecular simulations, it was shown that for model slit pores the capacitance oscillates as a function of the pore size.~\citep{Feng11c,Jiang11,Wu11} Such a phenomenon is nonetheless unlikely to be observed in real porous carbons with a diversity of pore sizes. Kondrat~et~al. have indeed shown through an analytical theory that polydispersity tends to smoothen features.~\citep{Kondrat12} Very recent molecular simulations works on regular tridimensional porous geometries have also shown that the pore size,~\citep{Liu19,Lahrar20} while being often used as a descriptor for the porous materials, is not necessarily well correlated with the structural, dynamical and capacitive properties of the systems.

In this work we conduct molecular dynamics simulations of model supercapacitors with two nanoporous carbon electrodes intentionally chosen so that they have the same average pore size but very different pore size distributions and morphologies. One of the carbon electrodes has a regular structure with a well defined pore size while the other is very disordered. We show that these characteristics lead to large differences in the capacitive properties: the ordered carbon has a much larger capacitance than the disordered one. We correlate these observations with the interfacial properties: i) the charge separation between anions and cations seems to be facilitated in the ordered carbons, related to higher confinements, and ii) in the disordered carbon, some carbon atoms are isolated from the liquid, and hence do not participate in the charge storage. Overall, these results confirm the importance of local morphology and suggest possible directions for screening porous carbons for optimized performance.

\section{Methods}

The model supercapacitors consist in two porous carbon electrodes in contact with neat [BMIM][PF$_6$] as the electrolyte. [BMIM][PF$_6$] is described by a coarse grained model with three sites for the cation and one site for the anion.~\citep{Roy10b} The cation geometry is kept rigid during the simulations. The intermolecular interactions are calculated as the sum of a Lennard-Jones potential and coulombic interactions: 
\begin{equation}
u_{ij}(r_{ij})=4\varepsilon_{ij} \left [ \left ( \frac{\sigma_{ij}}{r_{ij}}\right )^{12}- \left ( \frac{\sigma_{ij}}{r_{ij}}\right )^6 \right ]+\frac{q_iq_j}{4\pi\varepsilon_0r_{ij}}
\end{equation}
where $r_{ij}$ is the distance between sites $i$ and~$j$, $\varepsilon_0$ is the permittivity of free space, and $\sigma_{ij}$ and $\varepsilon_{ij}$ are the Lennard-Jones parameters. The parameters for the ions are taken from the work of Roy and Maroncelli~\citep{Roy10b} while those for the carbon atoms are taken from the article of Cole and Klein.~\citep{Cole83} Cross parameters are calculated using Lorentz-Berthelot mixing rules. The ions are considered non polarizable and carry a charge of $\pm0.78$~e. The scaling of the ion charge from $\pm1.0$~e to $\pm0.78$~e is used to improve the agreement between simulations and experiments for a variety of static and dynamic properties. The use of reduced charges is a computationally cheap approach to compensate for the use of a non polarizable force field. While its is obviously less accurate than the more computationally expensive polarizable force fields,~\citep{Schroder12} it was demonstrated to be a relevant approach for a large number of systems~\citep{Bowron10,Barghava07,Chaban11d}. The carbon structure is kept rigid as a single entity during the simulations. In reality, the structure shows a limited flexibility~\citep{Hantel11,Hantel12,Hantel14} but this aspect is difficult to integrate in the simulations conducted here which require the use of fluctuating charges on the electrodes. In a previous study on similar systems,~\citep{Lahrar20}  we have shown that a flexibility leading to relative height changes of 1\% and 2\% does not affect dynamical and structural properties significantly. 

In many works reporting molecular dynamics simulations of model supercapacitors, the systems are built by placing a dense liquid between two empty electrodes so that once the electrolyte has filled the porous materials, the density of the liquid in the center of the system is close to the experimental density. This method has the advantage that all the equilibration can be done in NVT simulations and no geometrical changes are needed but it can potentially lead to under-filled or over-filled pores if the amount of liquid or the distance between the electrodes are not estimated properly. It is indeed not straightforward to assess the quantity of liquid which will actually enter the pores, especially for ionic liquids where ionic interactions are very strong and can limit the accessibility to the pores. In this work, we have explored a new approach, which involves NPT simulations, to generate the starting configurations. This new methodology is described in Supporting information. The characteristics of the resulting model supercapacitors are given in Table~\ref{table:super-carac}.
\begin{table}[ht!]
\centering
\begin{tabular}{|c|c|c|c|c|c|}
\hline
Carbon & $L_x$ (\r{A}) & $L_y$ (\r{A}) & $L_z$ (\r{A}) & N$_{\rm C}$ & N$_{\rm pairs}$ \\
\hline
Ord & 48.44 & 48.44 & 270.51 & 10,998 & 1,373 \\ 
\hline
Disord & 43.55 & 43.55 & 277.80 & 9,138 & 1,274 \\
\hline
\end{tabular}
\caption{Characteristics of the model supercapacitors simulated in this work with [BMIM][PF$_6$] as the electrolyte. N$_{\rm C}$ and N$_{\rm pairs}$ are respectively the number of carbon atoms (electrodes and graphene sheets at the extremity of the box) and the number of ion pairs.}
\label{table:super-carac}
\end{table}

The model supercapacitors are simulated using the MetalWalls software.~\citep{MetalWalls}. This MD code allows for the simulation of supercapacitors and is adapted to the application of 2D periodic boundary conditions. The initial configurations were generated using the fftool software~\citep{fftool} and the NPT simulations used for the porous carbon filling were realised using LAMMPS.~\citep{LAMMPS} The rigidity of the cation is maintained using the SHAKE algorithm.~\citep{Ryckaert} After assembling the full model supercapacitor, simulations start with a constant charge equilibration step ($q~=~0$ for all electrode atoms) in the NVT ensemble for 2~ns. Then each system goes through a constant potential equilibration during which the electrode charge is monitored: the end of equilibration is defined when the charge reaches a plateau. The two potential differences studied in this work are $\Delta\psi$ = 0~V and 1~V. In previously reported works, different strategies have been explored to reach equilibrium,~\cite{Merlet12,Pean14,Liu19,Breitsprecher18,Fang19} including the application of a constant non zero charge but, while this is fast, it is hard to handle. Here, in order to reach faster equilibrium at 1~V, we adopted a new strategy which is described in Supporting information. 

\section{Model supercapacitors and their integral capacitances}

\begin{figure} [ht!]
\begin{center}
\includegraphics[scale=0.29]{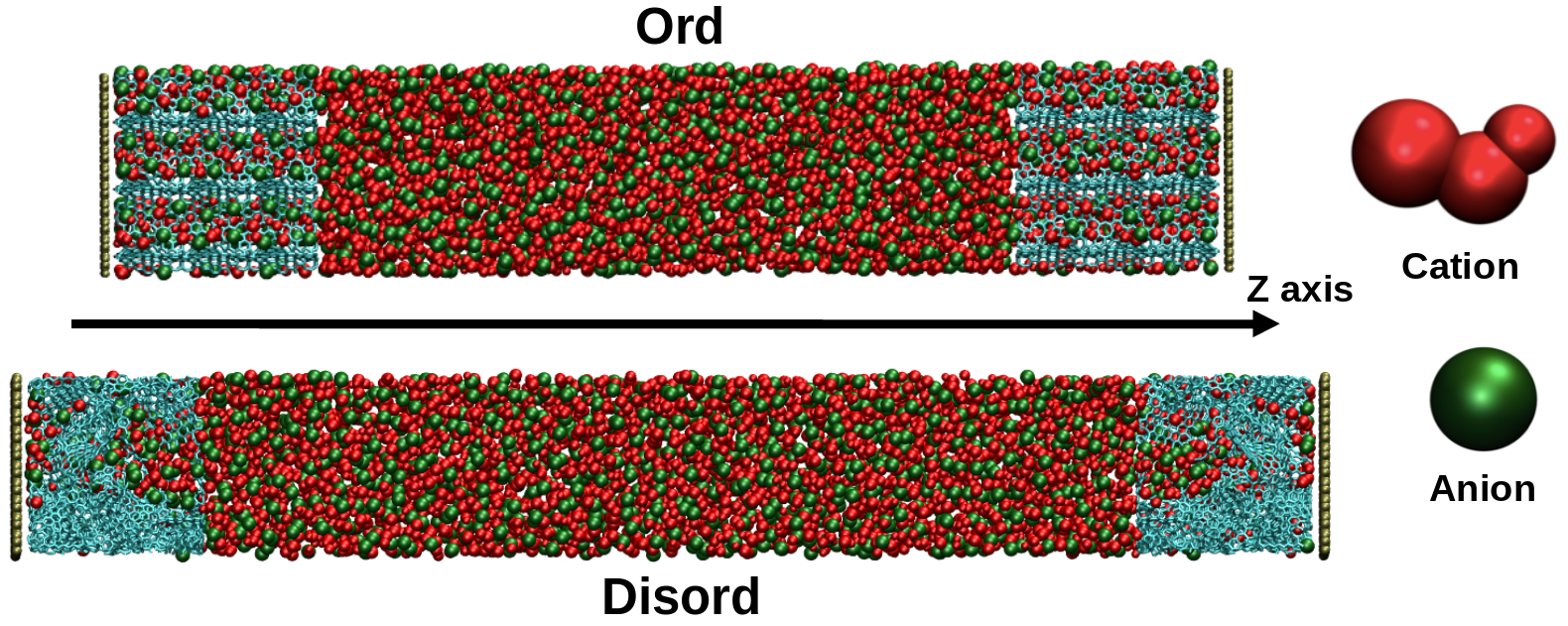}
\end{center}
\caption{Model supercapacitors simulated in this work: the anions are in green, the cations in red and the carbon atoms of the electrodes in blue. Atoms colored in brown are neutral carbon atoms used to prevent ions from leaving the simulation box (in these simulations there are no periodic boundary conditions in the $z$ direction to avoid unrealistic interactions between the positive and negative electrodes). The snapshot was generated using VMD.~\cite{VMD}}
\label{fig:model-supercap}
\end{figure}

Molecular dynamics simulations of model supercapacitors consisting in two porous carbon electrodes in contact with neat [BMIM][PF$_6$] as the electrolyte were performed. The simulations conducted here have some specificities in terms of building the initial configurations and equilibrating the systems at constant potential for which details are provided in Supporting information. For each studied system, the two electrode structures are perfectly identical. For one of the supercapacitors, the electrode structure is regular and designated as ``Ord" in the remainder of this article. The other supercapacitor is built with highly disordered electrodes and is designated as ``Disord". Both electrode structures have the same average pore size (12.2~\r{A}) and the same density (1~g~cm$^{-3}$). Figure~\ref{fig:model-supercap} illustrates the model supercapacitors studied in this work.

\begin{figure} [ht!]
\begin{center}
\includegraphics[scale=0.3]{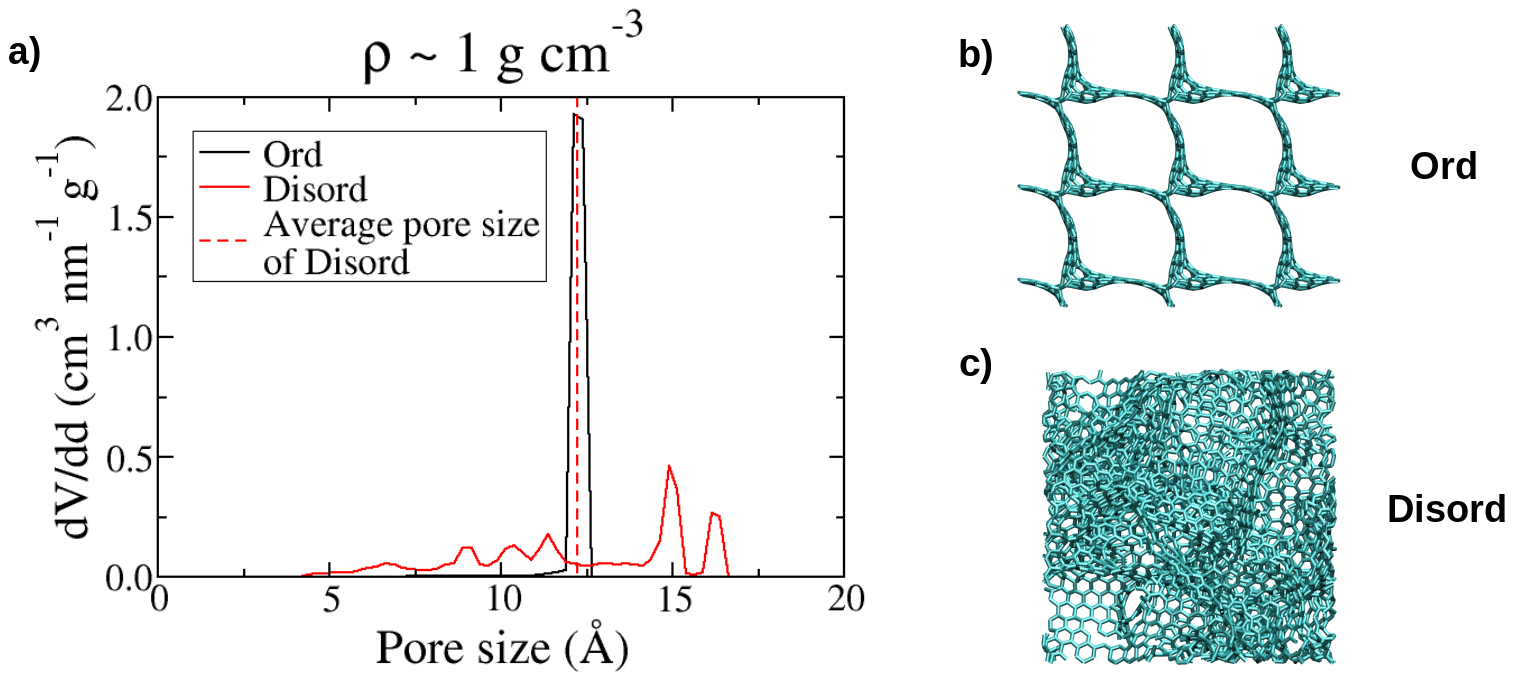}
\end{center}
\caption{a) Pore size distributions of the two porous carbon structures used in this work. b) The Ord carbon is characterized by a unimodal pore size distribution while c) the Disord carbon has a much wider pore size distribution. The pore size distributions were obtained using Poreblazer.~\citep{Sarkisov11}}
\label{fig:PSD-Structures}
\end{figure}

The two porous carbons studied were taken from published works~\citep{Deringer18,Palmer10} and have been generated through Quench Molecular Dynamics. The Disord structure is a relatively large porous carbon with 3,872 atoms generated using a classical force field. The Ord structure is originally a much smaller porous carbon with 172 atoms generated using a machine-learning based force field developed using the Gaussian Approximated Potentials approach.~\citep{Deringer17} The structure is not inherently regular but rendered so by replicating the initial coordinates two times in the three dimensions leading to a much more ordered structure with 4,644 atoms. Most importantly, the pore size distribution of the Ord carbon is very narrow and unimodal while the pore size distribution for the Disord carbon is much wider as shown in Figure~\ref{fig:PSD-Structures}. In terms of topology, the Ord carbon is characterized by tunnels along the $z$ axis while there are no preferred directions for the Disord carbon. The latter contains large pores but also smaller pores. 
The local structures of the porous materials studied are also similar, as visible from the carbon-carbon pair distribution functions which show the typical peaks of graphite like materials and rings statistics with a large proportion of 6-membered rings (see Supporting information). It is worth noting though that the peaks in the pair distributions functions of the Ord carbon are narrower than the ones of the Disord carbon, which is consistent with a more ordered structure.   

After simulating the model supercapacitors at 1~V, the first property we evaluate is the integral capacitance which corresponds to the amount of charge stored on the electrodes for a given potential difference. For a single electrode, the integral capacitance is defined as:
\begin{equation}
C_{int} = \frac{<Q_{tot}>_{\Delta\psi}}{\Delta\psi_{elec}}
\end{equation} 
where $<Q_{tot}>$ is the average total charge on the electrode under a constant potential $\Delta\psi$ and $\Delta\psi_{elec}$ is the potential drop between the electrode and the electrolyte. The average electrode charge is extracted at equilibrium from the curves shown in Supporting information. While it is not straightforward to calculate precisely the potential drop between each electrode and the electrolyte,~\cite{Pean15} it is possible to estimate that it is half of the 1~V imposed potential. Indeed, as the system is symmetrical and the ions have similar sizes, the capacitances of the two electrodes are expected to be similar.

The total electrode charges of the Ord and Disord systems as well as the corresponding capacitances are given in Table~\ref{table:capacitance}. To compare the two porous carbons, the capacitance values are normalized by the electrode mass as is often done for electrochemical characterization experiments. The capacitances can also be normalized by the accessible volumes or surfaces but these quantities are usually less well defined.~\cite{Stoeckli05,Stoeckli13}
\begin{table}[ht!]
\centering
\begin{tabular}{|c|c|c|}
\hline
System &  Electrode charge (e) & Capacitance (F.g$^{-1}$) \\
\hline
[BMIM][PF$_6$]/Ord & $\pm$ 38.6 ($\pm$ 0.5) & 134 \\ 
\hline
[BMIM][PF$_6$]/Disord & $\pm$ 21.2 ($\pm$ 0.5) & 88 \\ 
\hline
\end{tabular}
\caption{Total electrode charge and capacitance of one electrode for Ord and Disord systems under an applied potential difference of $\Delta\psi = $1~V.}
\label{table:capacitance}
\end{table}
According to these results, the ordered carbon has a better performance with an electrode capacitance equal to 134~F~g$^{-1}$ against 88~F~g$^{-1}$ for the disordered carbon, i.e. +~52\%. The values calculated values for the Disord carbon are close to the values observed experimentally, 90~F~g~$^{-1}$ and 75~F~g~$^{-1}$ for the negative and positive electrode
respectively, for the same ionic liquid in a CDC type carbon.~\cite{Pean15} The values for the Ord carbon are high compared to values reported previously for ordered zeolite templated carbons~\cite{Liu19} all in the range 60-100~F~g~$^{-1}$ (values are doubled compared to Figure~3 in that article due to the chosen way of calculating the capacitance). 

\section{Quantities of adsorbed ions and charge separation}

\begin{figure} [ht!]
\begin{center}
\includegraphics[scale=0.45]{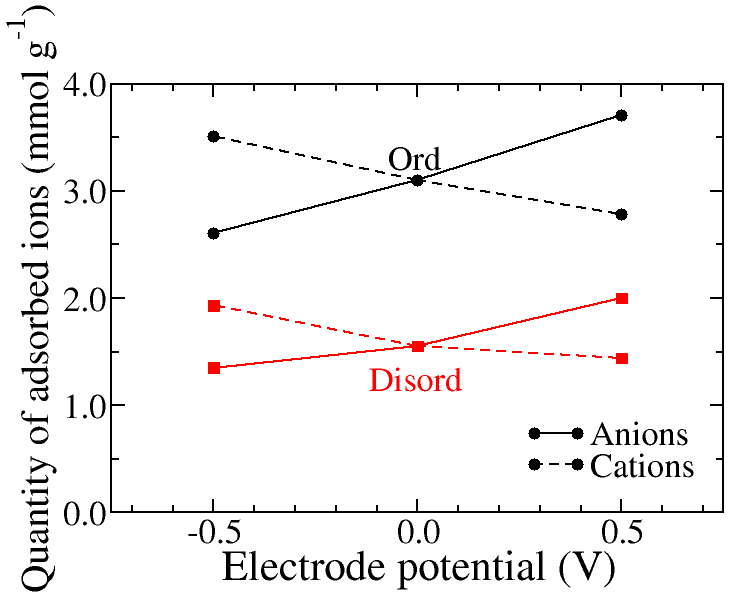}
\end{center}
\caption{Quantities of adsorbed ions in the Ord and Disord porous carbons as a function of the applied potential difference. }
\label{fig:Qty-ions}
\end{figure}

To understand the origin of this large difference in capacitance, we explore a number of properties starting with quantities of adsorbed ions. It has been shown on several occasions that, in accordance with a simple principle of electroneutrality, the charge carried by an electrode is equal to the opposite of the charge corresponding to the adsorbed ions.~\cite{Griffin15,Forse16} Figure~\ref{fig:Qty-ions} gives the quantities of adsorbed ions under 0~V and 1~V applied potential differences. The values reported here are consistent with previous experimental and molecular simulations studies,~\cite{Forse16} especially for the Disord carbon.
The quantity of adsorbed ions is approximately 1.9 times larger for the Ord system compared to the Disord one for all potentials. This is consistent with our previous study showing that total numbers of adsorbed ions were larger for a range of carbons with well defined pore sizes compared to disordered carbons of a similar type to the Disord structure studied here.~\cite{Lahrar20} The absolute ionic charge within the pores, $|N_{cations} - N_{anions}|$, is larger for the ordered structure, equal to 39.8~e ($\pm 0.8$), than for the disordered structure, equal to 20.7~e ($\pm 0.8$). The ionic charges are, as expected, very close to the electrode charges and within the error range. It is worth noting that the rate of change of the quantities of adsorbed ions with the applied potential, estimated from the slope of the roughly linear trends in Figure~\ref{fig:Qty-ions}, is also larger for the Ord carbon ($\pm$~0.9~mmol~g$^{-1}$~V$^{-1}$) compared to the Disord one ($\pm$~0.6~mmol~g$^{-1}$~V$^{-1}$). All these findings combined with the faster charging dynamics observed for the Ord system (see Supporting information) seem to indicate a better pore accessibility and charge separation for this system compared to the Disord one. 

The better accessibility can be due to a larger accessible volume for the Ord carbon. Indeed, while the densities of the two carbons are similar, their pore size distributions are very different and the spatial distribution of the carbon atoms can have a large impact on the accessible volume. The geometrical accessible volumes calculated using Poreblazer~\citep{Sarkisov11} are 75,418~\r{A}$^3$ for the Ord carbon and 47,428~\r{A}$^3$ for the Disord carbon, i.e. 1.6 times larger for the Ord carbon. This explains partly the 1.9 times larger quantity of adsorbed ions for the Ord carbon but is not sufficient to explain it fully. Using approximate volumes of 126~\r{A}$^3$ for the cation and 67~\r{A}$^3$ for the anion (evaluated from the Lennard-Jones parameters and cation geometry), the ionic liquid is estimated to fill around 45\% of the porous volume for the Ord carbon against only 31\% for the Disord carbon (see Supporting information for the values at various potentials). So not only is the geometric pore volume larger for the Ord carbon, the ionic liquid is also filling it more efficiently. It is worth noting that these percentages of pore filling are in the same range as the 40\% value obtained for two neat ionic liquids ([Pyr$_{13}$][TFSI] and [EMI][TFSI]) in the YP-50F activated carbon~\cite{Forse15} showing once again that molecular simulations are adapted to such studies. Overall, these analyses suggest that part of the porous volume and carbon atoms are not involved in charge storage in the Disord carbon. 

\section{Degrees of confinement of the electrolyte ions}

\begin{figure}[ht!]
\begin{center}
\includegraphics[scale=0.38]{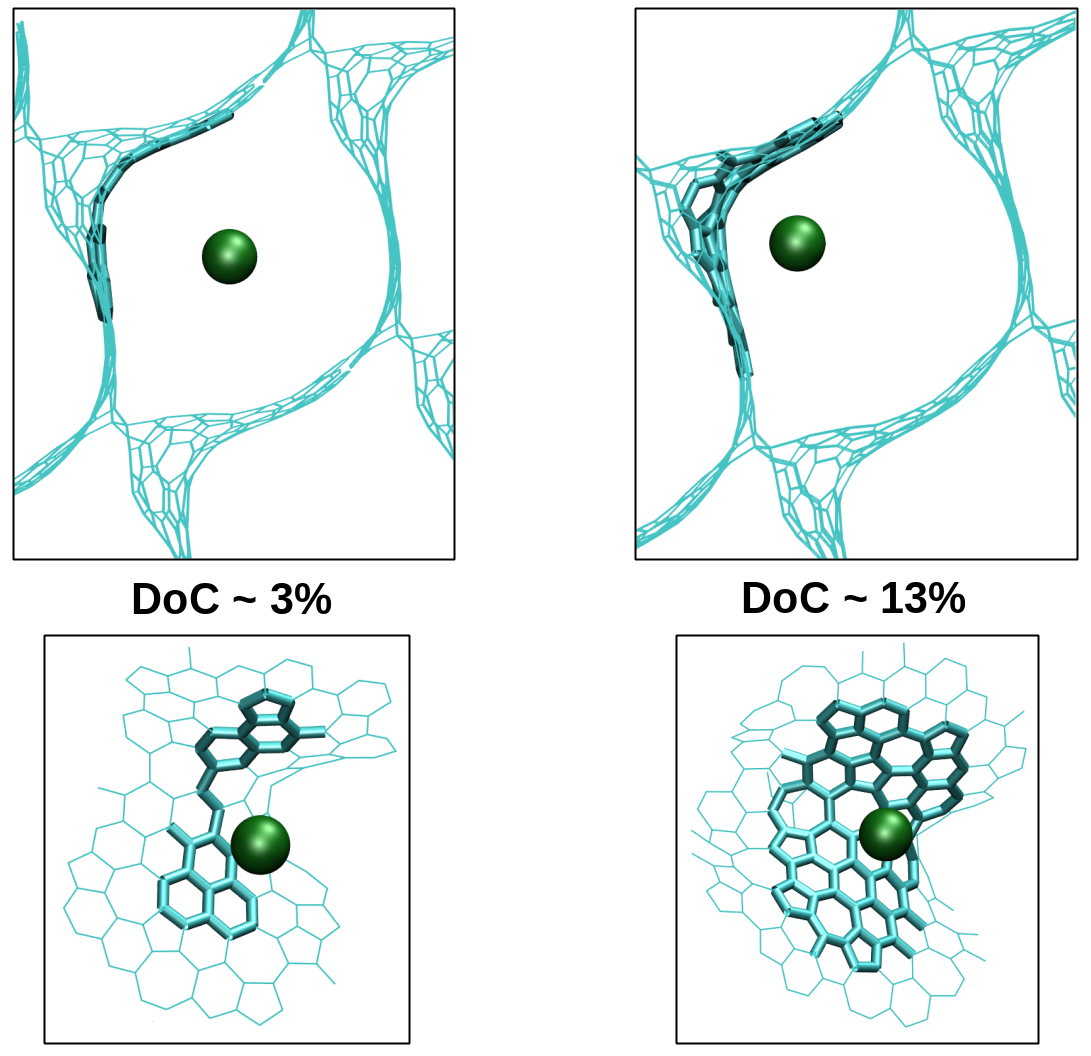}
\end{center}
\caption{Snapshots of anions (in green) in two different environments in the Ord structure. The carbon atoms are represented as solid lines along the carbon-carbon bonds with bolder lines for the carbon atoms belonging to the first coordination shell of the anions. Other anions and cations are not represented for simplicity. Small DoCs correspond to ions in the center of the pores while larger DoCs correspond to ions in the corners of the structure, i.e. in close contact with two surfaces.}
\label{fig:DoC}
\end{figure}

The charge carried by electrode atoms and the charge separation between anions and cations depend on both the applied potential and on the electrode-electrolyte interfacial structure. To characterize the behavior of the liquid near carbon atoms we calculate the degrees of confinement (DoCs) of the ions in the pores. This property was first defined in an attempt to analyze the correlation between confinement and charge storage efficiency~\cite{Merlet13d} and was later shown to be relevant in several systems to understand the capacitive properties of porous carbons.~\cite{Liu19,Prehal17} The degree of confinement is defined by the percentage of the solid angle around the ion, that is, in the first coordinating sphere, which is occupied by carbon atoms. The solid angle is normalized by the maximum value it can take so that values are always between 0\% and 100\%.~\cite{Merlet13d} Figure~\ref{fig:DoC} illustrates ions in different environments, more or less confined.

\begin{figure} [ht!]
\begin{center}
\includegraphics[scale=0.57]{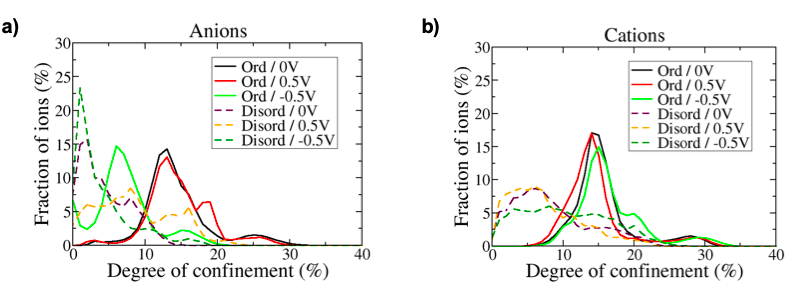}
\end{center}
\caption{Distributions of DoCs for the ions adsorbed in the electrodes of the Ord and Disord structures under potential differences of 0~V and 1~V. The cut-offs used to define the first coordination shell are given in Supporting information.}
\label{fig:DoC-Ord-Disord}
\end{figure}

Figure~\ref{fig:DoC-Ord-Disord} shows the distributions of DoCs of the confined ions in the two porous structures for different electrode potentials.
There are two striking features in this figure. Firstly, the distributions of DoCs are much broader for the Disord carbon compared to the Ord carbon which is expected as the variety of environments is reduced in the ordered structure. Indeed, it is consistent with Figure~\ref{fig:DoC} showing that, in the Ord structure, different DoCs basically correspond to larger distances to the carbon surface and not to different arrangements of the carbon atoms. Secondly, the DoCs are larger in the Ord carbon compared to the Disord carbon. This is qualitatively in agreement with a larger capacitance for the Ord carbon since ions closer to the surface will induce a larger charge on the carbon and will also increase the screening of the ion-ion interaction by the surface which will facilitate the charge separation.  

Taking the 0~V systems as a reference, it is found that, at the positive Ord electrode, the DoC of the anions remains almost the same except for the appearance of a bump towards a high degree of confinement of 19\%. This peak corresponds to the formation of a population of more confined anions but the little difference between the neutral and positively charged electrodes shows that most anions are already adsorbed on the carbon surface in the neutral electrode. For the Disord structure, the change in confinement is more remarkable with a large increase in the number of anions with a DoC between 10\% and 20\% which is almost zero at 0~V. From 0~V to -0.5~V in the Ord electrode, the DoC distributions of the anions show a shift towards the less important degrees of confinement without modifying the shape of the curve. This suggests that the remaining anions in the electrode do not necessarily change the type of adsorption site but move away from the surface. For the Disord carbon, the shape of the curve is modified with the anions getting less confined. For cations, the variations in the distributions of DoCs with the change in potential difference are present, but less marked than for anions. That is to say that although participating in the mechanisms of adsorption/desorption and therefore in charge storage, the local structure of the confined cations is less influenced by the applied potential.

\section{Electrode atoms charges and accessibility}

\begin{figure} [ht!]
\begin{center}
\includegraphics[scale=0.45]{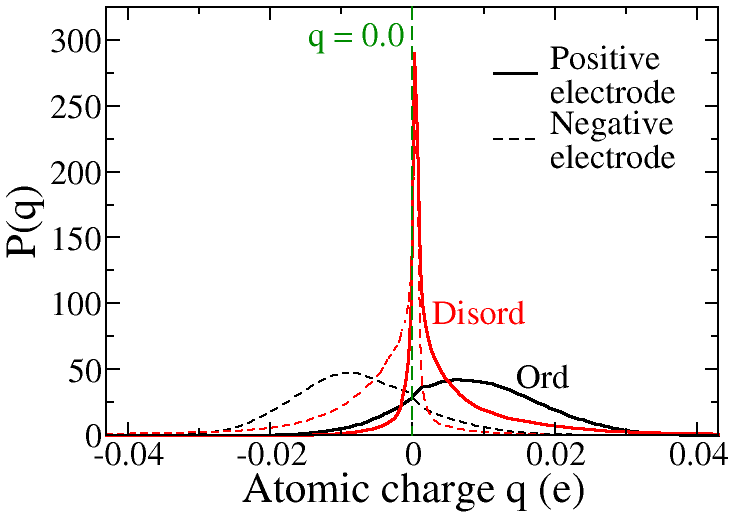}
\end{center}
\caption{Histograms of instantaneous charges on the electrode atoms for the 1~V potential difference. Histograms of atomic charges averaged over the entire trajectory are similar to the instantaneous ones and are thus not reported here.}
\label{fig:histo-charge}
\end{figure}

The quantities of adsorbed ions and the degrees of confinement suggest that some parts of the disordered carbon structure are inaccessible to the ions. We now turn to a more detailed analysis of the electrode side of the interface to confirm this hypothesis and assess the consequences of such a limited accessibility on the charge storage efficiency. Figure~\ref{fig:histo-charge} shows the probability densities of instantaneous charges on the electrode atoms for both porous carbons. It is interesting to note that the probability densities of atomic charges averaged over the entire trajectories, which are the ones considered for example by Liu~\emph{et al.}~\cite{Liu19}, are very similar and are thus not reported here. These quantities once again demonstrate the large difference between both systems. For the Disord structure, many carbon atoms have a zero or close to zero charge, meaning that they do not participate efficiently in the charge storage. For the Ord structure, on the contrary, there are many charges around $\pm~0.01$~e and only few charges near zero. Large numbers of atoms with a zero charge have also been observed in molecular simulations of zeolite templated carbons based supercapacitors~\cite{Liu19} but the tail at large atomic charges, also seen in previous works,~\cite{Merlet12} seems to be more specific of disordered porous carbons. The reduced number of carbon atoms with a low charge and the larger occurrence of carbon atoms with a large charge both make the charge storage more efficient in the Ord carbon. These differences could be related to local topologies which is what is investigated thereafter.

As the charges carried by the electrode atoms depend both on the applied potential and on the liquid nearby, it is expected that the charges with a value close to zero are located in regions of the carbon structure isolated from the liquid. In this work, a carbon atom is considered isolated if there is no ion at a distance less than $R_{cut}$ of its position. If there is at least one ion at a distance less than $R_{cut}$, then the carbon is not considered isolated. $R_{cut}$ is the distance corresponding to the limit of the first coordination shell between carbon atoms and anions or cations, i.e. the first minimum of the corresponding radial distribution function. 
\begin{figure}[ht!]
\begin{center}
\includegraphics[scale=0.37]{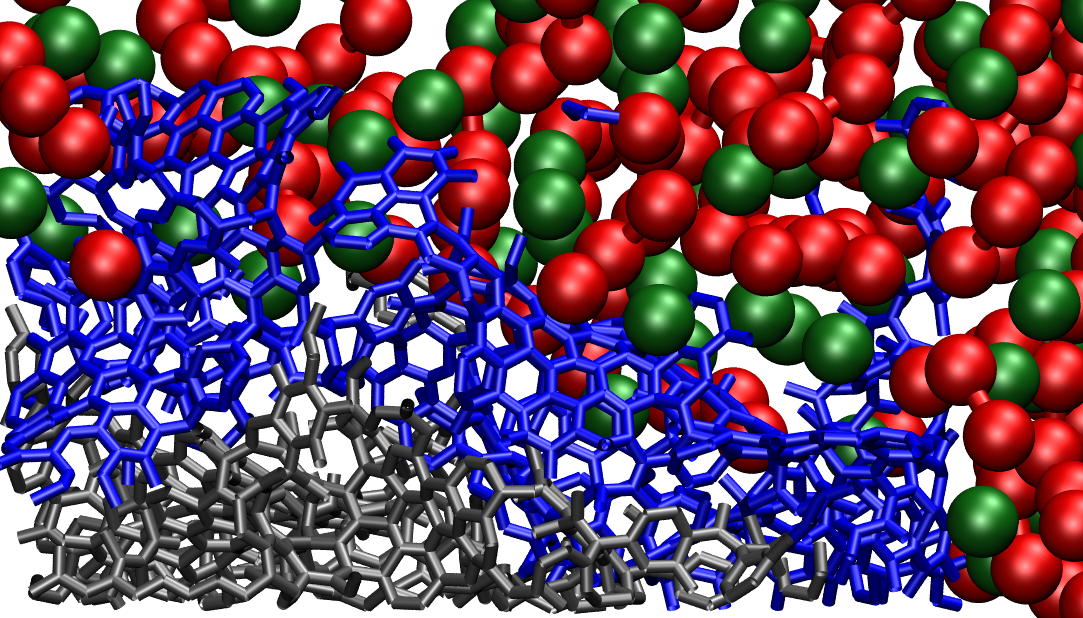}
\end{center}
\caption{Illustration of carbon atoms isolated from the electrolyte in the Disord system. The carbon atoms are represented in grey for isolated carbon atoms, located in a region inaccessible to the anions (in green) and cations (in red), and in blue for non isolated carbon atoms.}
\label{fig:isolated-example}
\end{figure}
An illustration of isolated carbon atoms is provided in Figure~\ref{fig:isolated-example}. Following this definition for isolated atoms, a carbon atom can be isolated throughout the duration of the trajectory, for a part of it or not at all. Table~\ref{tableau:isolated} gathers the number of carbons that are isolated for the entire trajectory for the two types of carbon. For the Ord carbon, all of the carbon atoms are in contact, at least part of the time, with ions. For the Disord carbon, on the contrary, a rather large proportion of the carbons ($>$~16\%) is never in contact with the ionic liquid. This is intuitively in agreement with a large number of carbons having a zero charge in the Disord-based supercapacitor.
\begin{table}[ht!]
\centering
\begin{tabular}{|c|c|c|}
\hline
Electrode & Ord & Disord \\
\hline
Negative & 2 ($\sim$ 0\%) & 924 (24\%) \\
\hline
Positive & 0 (0\%) & 620 (16\%) \\
\hline
\end{tabular}
\caption{Number of carbon atoms isolated from the ionic liquid during the whole duration of the trajectory for the two simulated systems. The corresponding percentages, relative to the total number of atoms in the electrode structure, are shown in parentheses.}
\label{tableau:isolated}
\end{table}

\begin{figure} [ht!]
\begin{center}
\includegraphics[scale=0.29]{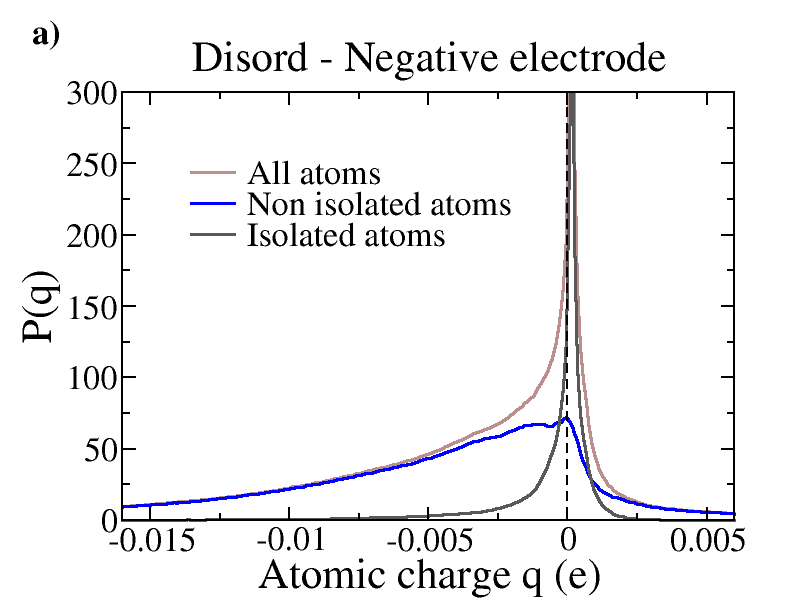}
\includegraphics[scale=0.29]{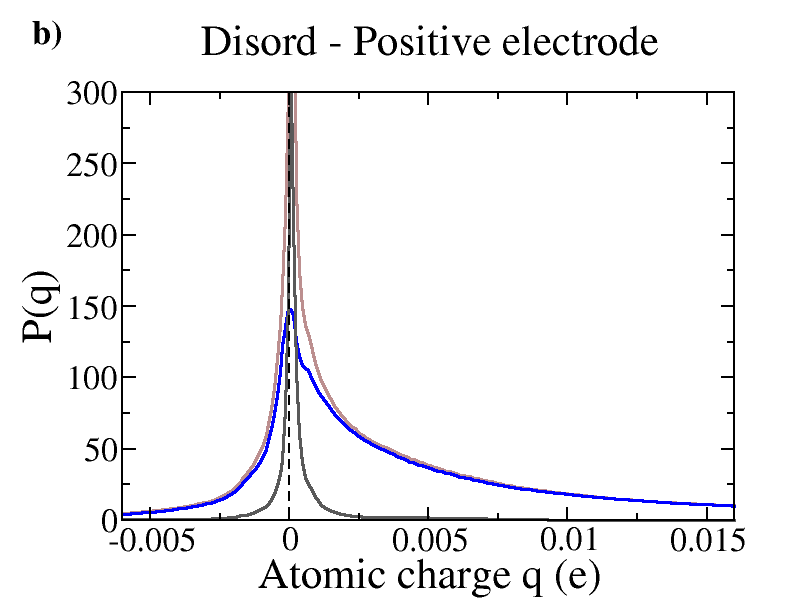}
\end{center}
\caption{Histograms of instantaneous charges for the negative (a) and positive (b) electrodes distinguishing between carbon atoms isolated or not from the adsorbed ions.}
\label{fig:histo-excluded}
\end{figure}

To check that isolated carbons indeed have a charge close to zero, histograms of electrode atom charges are plotted separately for carbon atoms isolated or not from the adsorbed electrolyte ions. The resulting curves are shown in Figure~\ref{fig:histo-excluded}.
The histograms for the carbon atoms with no ions nearby present a very sharp peak around zero confirming that isolated electrode atoms do not develop large charges and as such reduce the charge storage efficiency. Simply removing the isolated carbon atoms (neglecting any stability effect) could lead in the present case to an increase of capacitance from 88~F~g$^{-1}$ to 105-115~F~g$^{-1}$ (considering 620 or 924 isolated carbon atoms). The existence of such isolated carbon atoms, in addition to the lower occupancy of the porous volume could be sufficient to explain the capacitance values observed here for Disord and Ord carbons. It is worth noting that while isolated atoms do not develop large charges, some atoms which are not considered isolated have a close to zero charge. In fact, around 50~\% of the carbon atoms which have a charge below 0.001~e during the entire trajectory are not considered isolated all the time. 

\begin{figure} [ht!]
\begin{center}
\includegraphics[scale=0.3]{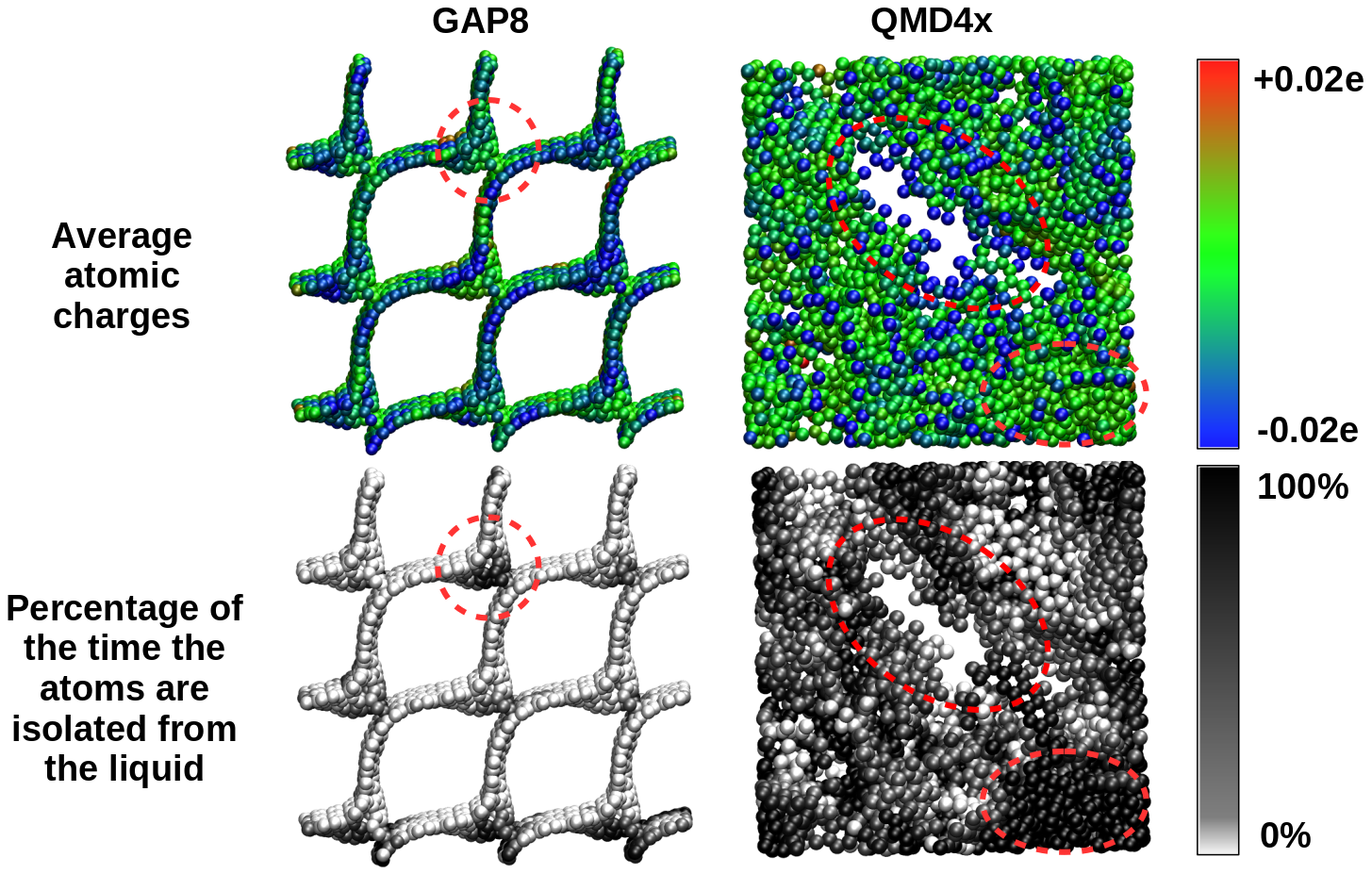}
\end{center}
\caption{Visualization of the average atomic charges and fraction of the trajectory during which the atoms are isolated from the ionic liquid for the negative electrodes in both systems at 1~V. Some areas are highlighted by dashed red circles and discussed in the text. These images were generated using VMD.~\cite{VMD}}
\label{fig:vis-carbons-neg}
\end{figure}

Using the molecular simulations results, it is also possible to visualize the electrode atom charges and how isolated these atoms are from the liquid directly on the atomistic carbon structures. Figures~\ref{fig:vis-carbons-neg} and~S6 present a visualization of the average electrode charges and the fraction of the trajectory during which the carbon atoms are isolated.
For the Ord carbon, the few zones where the carbon atoms are isolated for most of the simulation correspond to a ``funnel" region marked with a red dashed circle. The carbons present in these regions actually have charges relatively close to zero (colored in green). It should however be noted that the charges are far from being homogeneous in the rest of the structure, in agreement with the histograms of charges given in Figure~\ref{fig:histo-charge}. For the Disord carbon, it is more difficult to analyze the results. The carbon atoms surrounding the large pore present in the structure appear to be less isolated and with varying loads. On the contrary, there are zones where the carbons are isolated most of the time and these areas appear to correspond to atomic charges close to zero, in agreement with the charge histograms shown in Figure~\ref{fig:histo-excluded}. It is worth noting that, as evidenced in previous works,~\cite{Limmer13b} the charge on a given electrode atom does not depend only on the nearby electrolyte (and the actual ion-carbon distance) but also on the charge of neighboring electrode atoms so a direct link between the charge on a given carbon atom and the fraction of the time it is isolated is not expected to exist. Nevertheless, the analysis carried out here confirms that a porous structure with many inaccessible carbon atoms will probably show relatively poor performances in terms of capacitive properties. Since the quantity of adsorbed ions and maximal degrees of confinement do not change much with the applied potential difference, it might be possible to assess the proportion of isolated carbons already from simulations with neutral porous carbons, much less computationally expensive. If indeed correlated with the capacitance, this could serve as a screening process for assessing the electrochemical performance of a variety of carbons. This could be explored in the future using a larger set of carbons, for example those studied by Liu~\emph{et al.}~\cite{Liu19}, for which capacitances have been already calculated.

\section{Conclusion}

Molecular simulations of two model supercapacitors based on carbon electrodes with the same density and average pore size but different porous structures have been performed. The capacitance for the carbon with the more ordered structure was shown to be much larger (+~52\%) than the one for the more disordered structure. Several reasons are demonstrated to contribute to this higher capacitance for the ordered porous carbon. Firstly, the quantity of adsorbed ions initially in the pores at 0~V is higher for the Ord carbon than for the Disord carbon. Secondly, the electrolyte ions have higher degrees of confinement in the Ord carbon. These two facts can ease the charge separation in the Ord carbon leading to a larger ionic charge. Finally, the Disord carbon contains a relatively high proportion of carbon atoms which are far from the liquid and correspond to low induced charges. While all these factors can have an impact, it is hard to assess what is their relative importance as these are not independent from each other. 

The results obtained here, and in particular the potentially large impact of the initial quantity of adsorbed ions and of the proportion of electrode atoms inaccessible to the liquid, suggest potential avenues for computationally cheap methods to screen porous carbons for supercapacitor applications. In addition, reducing the number of isolated electrode atoms might be more easily achieved using relatively ordered structures which could be obtained experimentally through templating with regular precursors.

\begin{acknowledgements}
This project has received funding from the European Research Council (ERC) under the European Union's Horizon 2020 research and innovation program (grant agreement no. 714581). This work was granted access to the HPC resources of CALMIP supercomputing center under the allocations P17037.
\end{acknowledgements}

\section*{Supplementary material}
Pair  distribution  functions  and  ring  statistics  for  the porous structures, additional details on the generation of initial coordinates for the model supercapacitors, a description of the equilibration strategy, estimations of porosity filling at various potentials, cut-offs for the calculations of DoCs and a visualization of the isolated electrode atoms and their charges for positive electrodes are provided in Supporting information.

\section*{Data availability}

The data corresponding to the plots reported in this paper, as well as example input files for MetalWalls, are openly available in the Zenodo repository with identifier 00.0000/zenodo.0000000.

\end{document}